\documentclass[pra,twocolumn,superscriptaddress,showpacs,preprintnumbers,amsmath,amssymb]{revtex4}
\usepackage{}

\usepackage{bm}% bold math
\usepackage{bbm}
\usepackage{amssymb}
\usepackage{amsfonts}
\usepackage{epsfig,graphicx}
\usepackage{amstext}
\usepackage{amsmath}
\usepackage{graphicx}
\usepackage{times}
\usepackage{txfonts}
\usepackage{dcolumn}
\usepackage{color}
\usepackage{dsfont}

\newcommand{\iden}{\mathds{1}}
\newcommand{\tr}{\mathrm{Tr}}

\newcommand{\etal}{\textit{et al. }}

\begin{document}
%%%%%%%%%%%%%%%%%%%%%%%%%%%%%%%%%%%%%%%%%%%%%%%%%%%%%%%%%%%%%%%%%%%%%%
%TCIDATA{OutputFilter=Latex.dll}
%TCIDATA{Version=5.00.0.2552}
%TCIDATA{<META NAME="SaveForMode" CONTENT="1">}
%TCIDATA{LastRevised=Wednesday, June 22, 2005 16:21:09}
%TCIDATA{<META NAME="GraphicsSave" CONTENT="32">}

\title{Limits on sequential sharing of nonlocal advantage of quantum coherence}

\author{Ming-Liang Hu}
\email{mingliang0301@163.com}
\affiliation{School of Science, Xi'an University of Posts and Telecommunications, Xi'an 710121, China}

\author{Jia-Ru Wang}
\affiliation{School of Science, Xi'an University of Posts and Telecommunications, Xi'an 710121, China}

\author{Heng Fan}
\email{hfan@iphy.ac.cn}
\affiliation{Institute of Physics, Chinese Academy of Sciences, Beijing 100190, China}
\affiliation{CAS Center for Excellence in Topological Quantum Computation, University of Chinese Academy of Sciences, Beijing 100190, China}
\affiliation{Songshan Lake Materials Laboratory, Dongguan 523808, China}

\begin{abstract}
%%%%%%%%%%%%%%%%%%%%%%%%%%%%%%%%%%%%%%%%%%%%%%%%%%%%%%%%%%%%%%%%%%%%%
Sequential sharing of nonlocal correlation is inherently related
to its application. We address the question as to how many observers
can share the nonlocal advantage of quantum coherence (NAQC) in a
$(d\times d)$-dimensional state, where $d$ is a prime or a power of a prime.
We first analyze the trade-off between disturbance and information
gain of the general $d$-dimensional unsharp measurements. Then in a
scenario where multiple Alices perform unsharp measurements on one
party of the state sequentially and independently and a single Bob
measures coherence of the conditional states on the other party, we
show that at most one Alice can demonstrate NAQC with Bob. This limit
holds even when considering the weak measurements with optimal pointer
states. These results may shed light on the interplay between nonlocal
correlations and quantum measurements on high-dimensional systems and
the hierarchy of different quantum correlations.
\end{abstract}


\maketitle

\section{Introduction} \label{sec:1}
%%%%%%%%%%%%%%%%%%%%%%%%%%%%%%%%%%%%%%%%%%%%%%%%%%%%%%%%%%%%%%%%%%%%%
The characteristics of a quantum state without any classical analog
are fundamental and key issue of quantum physics \cite{Nielsen}.
Formally, one can introduce different forms of nonlocal correlations
to characterize these intriguing characteristics, including Bell
nonlocality confirming the nonexistence of the local hidden variable
model \cite{Bell1,Bell2}, Einstein-Podolsky-Rosen (EPR) steering
confirming the nonexistence of the local hidden state model
\cite{steer1,steer2}, quantum entanglement originating from the
superposition principle of states \cite{QE}, and quantum discord
which is rooted in the noncommutativity of operators \cite{QD}.
These nonlocal correlations are crucial physical resources for
quantum communication and computation tasks which outperform their
classical counterparts.

For a given quantum state, when one assumes no-signaling among its
parties, the monogamy relation imposes constraints on the number
of observers who can share the quantum correlations in this state
\cite{monoe,monon,monos,monoc,monod}. But if the no-signaling
condition is partially relaxed, e.g.,
a single Bob holds half of an entangled pair, while multiple
Alices (say, Alice$_1$, Alice$_2$, etc.) hold the other half of that
pair and perform weak measurements sequentially and independently
on their half, then the prior measurement of Alice$_1$ implicitly
signals to Alice$_2$ by her choice of measurement setting,
likewise, Alice$_2$ signals to Alice$_3$, and so on.
Thereby the monogamy constraints might be relaxed to allow sequential
sharing of quantum correlations. In this context, a double
violation of the Clauser-Horne-Shimony-Holt (CHSH) inequality with
equal sharpness of measurements has been theoretically predicted
\cite{shareBT1,shareBT2,shareBT3,shareBT4} and experimentally
verified \cite{shareBE1,shareBE2,shareBE3}. Further studies showed
that an unbounded number of CHSH violations can be
achieved using weak measurements with unequal sharpness
\cite{shareBT6,shareBT7,shareBT8}. The weak measurement scenario has
also been extended to investigate sequential sharing of tripartite
Bell nonlocality \cite{shareBT9}, EPR steering \cite{shareST1,
shareST2,shareST3,shareST4}, and bipartite entanglement \cite{shareET1}.
Then an open question to ask is whether other forms of quantum
correlations could be sequentially shared, especially for those
beyond the two-qubit case or stronger than Bell nonlocality.

As a fundamental property in quantum theory, quantum coherence is
not only an embodiment of the superposition principle of states,
but is also intimately related to quantum correlations among the
constituents of a system \cite{Ficek}. In particular, following the
resource theory of entanglement \cite{QE} and quantum discord \cite{QD},
Baumgratz \etal \cite{coher} introduced a resource theoretic
framework for quantifying coherence, within which the fascinating
properties and potential applications of coherence have been investigated
in a number of contexts \cite{Plenio,Hu,Wu,ad1,ad2,ad3}. The resource theory of
coherence can also be used to interpret those already known quantum
correlations \cite{coen1,coen2,coen3,coqd1,coqd2,coqd3,coqd4} and
introduce other quantifiers of correlations \cite{Hu}, among which is
the nonlocal advantage of quantum coherence (NAQC) \cite{naqc1,naqc2}.
It is captured by violations of the coherence steering inequalities,
which is similar to the Bell nonlocality captured by violation of
the CHSH inequality \cite{CHSH}. Specifically, for a
$(d\times d)$-dimensional state $\rho_{AB}$ with $d$ being a power of a prime
(hereafter, we call it the two-qudit state), when the steered
coherence on $B$ after local measurements on $A$ exceeds a threshold,
we say that there is NAQC in the sense that such a coherence is
unattainable for any product state. As for its hierarchy with other
quantum correlations, the set of states with NAQC forms a subset of
entangled states \cite{naqc1,naqc2} and for the $d=2$ case, it is
also a subset of Bell nonlocal states \cite{naqc3}.

In this work, we investigate how many Alices could sequentially
demonstrate NAQC with a single Bob. We first analyze the
information-disturbance trade-off of the $d$-dimensional unsharp
measurements. Then in the context of unsharp measurements, we show
that different from the sequential sharing of Bell nonlocality
\cite{shareBT1,shareBT2,shareBT3,shareBT4,shareBE1,shareBE2,
shareBE3,shareBT6,shareBT7,shareBT8,shareBT9}, EPR steering
\cite{shareST1,shareST2,shareST3,shareST4}, and entanglement
\cite{shareET1}, at most one Alice can demonstrate the NAQC with Bob.
In particular, such a limit exists even when one considers the
weak measurements with optimal pointer states. These results may
enrich our comprehension on the interplay between NAQC and
measurements on high-dimensional systems.

\section{Characterization of the NAQC} \label{sec:2}
%%%%%%%%%%%%%%%%%%%%%%%%%%%%%%%%%%%%%%%%%%%%%%%%%%%%%%%%%%%%%%%%%%%%%
In 2014, Baumgratz \textit{et al.} \cite{coher} introduced a resource
theoretic framework for quantifying coherence. Within this framework,
the incoherent states are defined as those described by the diagonal
density operators, and for a given state described by the density
operator $\rho$, the amount of coherence could be quantified by its
minimal distance to the set $\mathcal{I}$ of incoherent states in the
same Hilbert space. By fixing the basis $\{|i\rangle\}$ which can be
recognized as the normalized eigenbasis of a Hermitian operator
$\mathcal{O}$, Baumgratz \textit{et al.} \cite{coher} further
identified two coherence measures, that is, the $l_1$ norm and
relative entropy of coherence, which are given by
%%%%%%%%%%%%%%%%%%%%%%%%%%%
\begin{equation}\label{eq2-1}
 C_{l_1}^\mathcal{O}(\rho)= \sum_{i\neq j} |\langle i| \rho|j\rangle|,~
 C_{re}^\mathcal{O}(\rho)= S(\rho_{\mathrm{diag}})-S(\rho),~
\end{equation}
%%%%%%%%%%%%%%%%%%%%%%%%%%%
where the subscripts $l_1$ and $re$ indicate the metrics of the two
coherence measures, while $S(\rho)$ and $S(\rho_{\mathrm{diag}})$
are the von Neumann entropies of $\rho$ and $\rho_{\mathrm{diag}}=
\sum_i \langle i |\rho |i\rangle |i\rangle\langle i|$, respectively.

Starting from the above coherence measures, one can then introduce
the NAQC which captures the nonlocal property of a bipartite state.
Specifically, the NAQC characterizes the ability of one party to
steer the coherence of the other one when they share a two-qudit
state $\rho_{AB}$. To illustrate such a nonlocal characteristics, we
suppose qudit $A$ ($B$) belongs to Alice (Bob) and denote by
$\{A^v\}$ the set of $d+1$ mutually unbiased observables. Alice
measures randomly one of the observables on qudit $A$ and informs
Bob of her choice $A^v$ and outcome $a$. Then the conditional state
of qudit $B$ will be given by
%%%%%%%%%%%%%%%%%%%%%%%%%%%
\begin{equation}\label{eq2-2}
 \rho_{B|\Pi^v_a}= \tr_A[(\Pi^v_a \otimes \iden)\rho_{AB}(\Pi^v_a \otimes \iden)]/p_{\Pi^v_a},
\end{equation}
%%%%%%%%%%%%%%%%%%%%%%%%%%%
where $\Pi^v= \{\Pi^v_a\}$ denotes the measurement operator of Alice,
$\iden$ is the identity operator, and $p_{\Pi^v_a}=
\tr[(\Pi^v_a \otimes \iden) \rho_{AB}]$ is the probability of Alice's
measurement outcome $a$.

After Alice's local measurements, Bob can measure coherence of the
conditional states on $B$ in different reference bases. In the first
framework, Bob chooses with equal probability $1/d$ one of the
eigenbasis of $\{A^u\}_{u\neq v}$, then one can obtain the average
steered coherence (ASC) $N_{AB}^\alpha(\rho_{AB})$ ($\alpha=l_1$ or
$re$) and the criterion for capturing NAQC in $\rho_{AB}$ is given
by \cite{naqc1,naqc2}
%%%%%%%%%%%%%%%%%%%%%%%%%%%%%
\begin{equation}\label{eq2-3}
 N_{AB}^\alpha(\rho_{AB})= \frac{1}{d}\sum_{u\neq v,a} p_{\Pi^v_a}
                           C_\alpha^{A^u}(\rho_{B|\Pi^v_a})
                         > N_c^\alpha,
\end{equation}
%%%%%%%%%%%%%%%%%%%%%%%%%%%%%
where the critical value $N_c^\alpha$ is obtained by first summing
the single-qudit coherence over the $d+1$ mutually unbiased bases and
then maximizing it over all the single-qudit states \cite{naqc2}. In
the second framework, Bob chooses the eigenbasis of $A^{\beta_v}$
after Alice's measurement $\Pi^v$, with $\beta=\{\beta_v\}_{v=0}^d$
being a permutation of the set $\{0,1,\ldots,d\}$ with elements
$\beta_v$. Then the criterion for capturing NAQC in $\rho_{AB}$
becomes \cite{naqc2}
%%%%%%%%%%%%%%%%%%%%%%%%%%%%%%
\begin{equation} \label{eq2-4}
 \mathcal{N}_{AB}^{\alpha}(\rho_{AB})= \max_{\{\beta_v\}} \sum_{v,a}
                                       p_{\Pi^v_a} C_{\alpha}^{A^{\beta_v}}(\rho_{B|\Pi^v_a})
                                     > N_c^\alpha,
\end{equation}
%%%%%%%%%%%%%%%%%%%%%%%%%%%%%%
where the maximum is taken over the $(d+1)!$ (the factorial of $d+1$)
possible permutations of $\{0,1,\ldots,d\}$.

For the special $d=2$ case, the sharing of NAQC captured by the
criterion of Eq. \eqref{eq2-3} has been studied
\cite{sharenaqc,sharenaqce}. However, Eq. \eqref{eq2-3} is less
efficient than Eq. \eqref{eq2-4} in capturing NAQC \cite{naqc2},
hence we will focus on the latter when discussing sharing of NAQC
by sequential observers.

\section{Framework of unsharp measurements} \label{sec:3}
%%%%%%%%%%%%%%%%%%%%%%%%%%%%%%%%%%%%%%%%%%%%%%%%%%%%%%%%%%%%%%%%%%
In the framework of von Neumann-type measurement \cite{von}, the
measurement process on a $d$-dimensional state $\rho_0$ implies
interaction of the system with the apparatus which induces the map:
$\mathcal {E}(\rho_0\otimes|\phi\rangle\langle\phi|) = \sum_{i j}
\Pi_i\rho_0 \Pi_j \otimes |\phi_i\rangle\langle\phi_j|$, where
$\{\Pi_i\}$ denotes the measurement operators, $|\phi\rangle$
is the initial pointer state of the apparatus, and
$|\phi_i\rangle$ is the postmeasurement state of the pointer
associated with the outcome $i$. By tracing out the pointer
states one can obtain the nonselective postmeasurement state as
$\rho= \sum_{ij}\Pi_i\rho_0 \Pi_j\langle\phi_i|\phi_j\rangle$,
where $\langle \phi_i |\phi_j \rangle$ may be different for
different $i\neq j$. Without loss of generality, here we consider
$\langle\phi_i|\phi_j\rangle\equiv F$ ($\forall i\neq j$) for
simplicity, then $\rho$ can be reformulated as
%%%%%%%%%%%%%%%%%%%%%%%%%%%%%
\begin{equation}\label{eq3-m1}
 \rho = F\rho_0+(1-F)\sum_i \Pi_i\rho_0\Pi_i,
\end{equation}
%%%%%%%%%%%%%%%%%%%%%%%%%%%%%
where $F\in[0,1]$ is the quality factor of the measurement, with
$F= 0$ corresponding to the usual projective (strong) measurement.
$F$ measures the extent to which the system state remains
undisturbed after the measurement and depends on the pointer of
the apparatus by its definition \cite{shareBT1}.

One point to be stressed here is that the reduced disturbance of
a weak measurement will induce reduced information gain. To
quantify such a quantity (i.e., the information gain or precision
of the measurements), one needs to choose a complete orthogonal
set of states $\{|\varphi_i\rangle\}$ as reading states because
the set $\{|\phi_i\rangle\}$ is non-orthogonal. As a result, the
probability of getting the outcome $i$ and the associated
(unnormalized) postmeasurement state $\rho_i=\langle \varphi_i|
\mathcal{E}(\rho_0\otimes|\phi\rangle\langle\phi|)|\varphi_i\rangle$
are given by
%%%%%%%%%%%%%%%%%%%%%%%%%%%%%
\begin{equation}\label{eq3-m2}
\begin{aligned}
 p_i= \, & \tr(\Pi_i\rho_0)|\langle\varphi_i|\phi_i\rangle|^2+ \sum_{j\neq i}\tr(\Pi_j\rho_0)|\langle\varphi_i|\phi_j\rangle|^2, \\
 \rho_i= \, & \Pi_i\rho_0\Pi_i |\langle\varphi_i|\phi_i\rangle|^2
              + \sum_{j\neq i} \Pi_j\rho_0\Pi_j |\langle\varphi_i|\phi_j\rangle|^2 \\
            & +\sum_{m\neq n} \Pi_m\rho_0\Pi_n \langle\varphi_i|\phi_m\rangle \langle\phi_n|\varphi_i\rangle,
\end{aligned}
\end{equation}
%%%%%%%%%%%%%%%%%%%%%%%%%%%%%
where $|\langle\varphi_i|\phi_i\rangle|^2$
($|\langle\varphi_i|\phi_{j\neq i}\rangle|^2$) is the probability
of obtaining the correct (wrong) outcome. When the measurement is
unbiased, i.e., $|\langle\varphi_i|\phi_i\rangle|^2$ is
independent of $i$ and $|\langle\varphi_i|\phi_{j\neq i}\rangle|^2$
is independent of $j\neq i$, Eq. \eqref{eq3-m2} can be
reformulated as
%%%%%%%%%%%%%%%%%%%%%%%%%%%%%
\begin{equation}\label{eq3-m3}
\begin{aligned}
 p_i= \, & G\tr(\Pi_i\rho_0)+\frac{1-G}{d}, \\
 \rho_i= \, & \frac{\mathcal{F}}{d}\rho_0+ \frac{1+d_1G- \mathcal{F}}{d}\Pi_i\rho_0\Pi_i \\
            & +\frac{1-G- \mathcal{F}}{d}\left(\sum_{j\neq i}\Pi_j\rho_0\Pi_j+\sum_{m\neq n \atop m,n\neq i}\Pi_m\rho_0\Pi_n \right),
\end{aligned}
\end{equation}
%%%%%%%%%%%%%%%%%%%%%%%%%%%%%
where $G=1-d |\langle\varphi_i|\phi_{j\neq i}\rangle|^2$ is the
precision of the measurement which quantifies the information gain
from the measured system and we have denoted by $\mathcal{F}=
[(1+d_1G)(1-G)]^{1/2}$, where $d_1=d-1$. $G$ also depends on the
pointer states of the measuring apparatus. Usually, one has the
trade-off $F^2+G^2\leqslant1$, and for $F^2+G^2=1$, the trade-off
is said to be optimal in the sense that the measurement yields the
highest precision for a given quality factor \cite{shareBT1}.

In this paper, we follow the framework of Refs. \cite{shareBT1,
shareBT2,shareST2} and consider the $d$-dimensional unsharp
measurements represented by the set of effect operators
%%%%%%%%%%%%%%%%%%%%%%%%%%%%%
\begin{equation}\label{eq3-1}
 E^v = \bigg\{E^v_a \mid E^v_a= \lambda \Pi^v_a+\frac{1-\lambda}{d}\iden, a=0,1,\ldots,d-1 \bigg\},
\end{equation}
%%%%%%%%%%%%%%%%%%%%%%%%%%%%%
where $\{E^v_a\}$ represents the measurement settings with $d$
possible outcomes per setting, $0< \lambda\leqslant 1$ represents the
sharpness parameter, and $\Pi^v_a=|\phi^v_a\rangle\langle\phi^v_a|$
is the projector. In the following, we restrict ourselves to
$\Pi^v_a$ ($v=0,1,\ldots,d$) constructed
by the $d+1$ mutually unbiased bases \cite{MUB1,MUB2}:
%%%%%%%%%%%%%%%%%%%%%%%%%%%%%
\begin{equation}\label{eq3-2}
\begin{aligned}
 &\left|\phi^0_a\right\rangle= \sum_{n=0}^{d-1} \delta_{an} |n\rangle,~
  \left|\phi^d_a\right\rangle= \frac{1}{\sqrt{d}} \sum_{n=0}^{d-1}
                               e^{i\frac{2\pi}{d}an} |n\rangle, \\
 &\left|\phi^r_a\right\rangle= \frac{1}{\sqrt{d}} \sum_{n=0}^{d-1}
                               e^{i\frac{2\pi}{d}r(a+n)^2} |n\rangle~ (r=1,\ldots,d-1),
\end{aligned}
\end{equation}
%%%%%%%%%%%%%%%%%%%%%%%%%%%%%
where $\delta_{an}$ is the Delta function and $i$ represents the
imaginary unit. One can note that the unsharp measurement operator
$E^v_a$ corresponds to a linear combination of the projector
$\Pi^v_a$ with the white noise. It satisfies the relation
$\sum_a E^v_a= \iden$ ($\forall v$) and belongs to the class of
positive-operator-valued measurements. In contrast to the
conventional strong measurement which enables an extraction of the
maximum information and destroys completely the system to be
measured, for the unsharp measurements the system is weakly coupled
to the probe and thus provides less information about the system
while producing less disturbance \cite{weak1,weak2}. Hence, the
postmeasurement states retain some original properties of the
measured system which might be observed by the subsequent observers.
Moreover, $E^v_a$ reduces to the projective (strong)
measurements when $\lambda=1$, and for such a special case, the
basis comprising $\Pi^v_a$ is an essential ingredient for
introducing the flag additivity condition which is equivalent to
the strong monotonicity and convexity of a coherence measure
\cite{new1,new2}.

Note that for $d=2$, the effect operators can also be written as
$E_\pm=(\iden\pm \lambda\hat{n}\cdot\vec{\sigma})/2$, where
$\hat{n}$ is a unit vector in $\mathbb{R}^3$ and $\vec{\sigma}$
is a vector composed of the three Pauli operators.

From Eq. \eqref{eq3-1} one can obtain that for any initial state
$\rho_0$, the nonselective postmeasurement state is given by \cite{Luders}
%%%%%%%%%%%%%%%%%%%%%%%%%%%%%
\begin{equation}\label{eq3-n1}
 \rho= \sum_a \sqrt{E^v_a} \rho_0 \sqrt{E^v_a}
     = \lambda_0 \rho_0 +(1-\lambda_0)\sum_a \Pi^v_a\rho_0\Pi^v_a,
\end{equation}
%%%%%%%%%%%%%%%%%%%%%%%%%%%%%
and the probability of getting the outcome $a$ is given by
%%%%%%%%%%%%%%%%%%%%%%%%%%%%%
\begin{equation}\label{eq3-n2}
 p_{E^v_a}= \tr(E^v_a \rho_0)= \lambda \tr(\Pi^v_a \rho_0)+ \frac{1-\lambda}{d},
\end{equation}
%%%%%%%%%%%%%%%%%%%%%%%%%%%%%
where
%%%%%%%%%%%%%%%%%%%%%%%%%%%%%
\begin{equation}\label{eq3-n3}
 \lambda_0= \frac{1}{d}\left[(d-2)(1-\lambda)
            +2\sqrt{1+(d-2)\lambda-d_1\lambda^2}\right].
\end{equation}
%%%%%%%%%%%%%%%%%%%%%%%%%%%%%
Then by comparing Eqs. \eqref{eq3-m1} and \eqref{eq3-m3} with Eqs.
\eqref{eq3-n1} and \eqref{eq3-n2}, one can see that the quality
factor and precision of the unsharp measurements \eqref{eq3-1} are
respectively given by
%%%%%%%%%%%%%%%%%%%%%%%%%%%%%
\begin{equation}\label{eq3-n4}
 F= \lambda_0,~ G= \lambda.
\end{equation}
%%%%%%%%%%%%%%%%%%%%%%%%%%%%%
Hence unsharpening the measurements with a parameter $\lambda$
enables the control of the trade-off between disturbance and
information gain. For $d=2$, one always has the optimal trade-off
$F^2+G^2=1$. But for the prime $d\geqslant 3$, $F^2+G^2\leqslant 1$,
and the equality holds only for $\lambda=1$, which corresponds to
$F=0$ and $G=1$, namely, the case of a projective (strong)
measurement \cite{shareBT1}.

\begin{figure}
\centering
\resizebox{0.44 \textwidth}{!}{%
\includegraphics{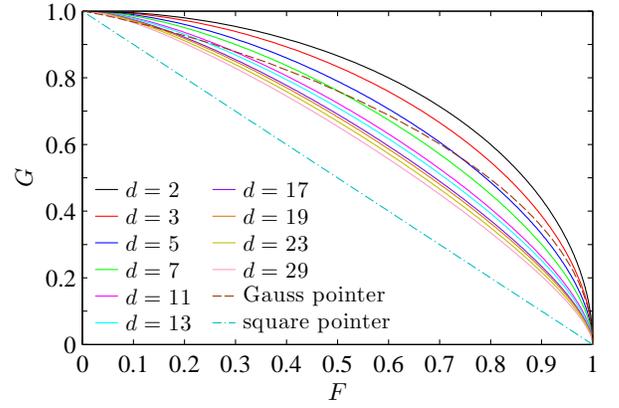}}
% If not, use\vspace{5cm}
% Give the correct figure height in cm
\caption{The trade-off between the quality factor $F$ and
precision $G$ for the unsharp measurements \eqref{eq3-1}. The
solid lines from top to bottom correspond to the primes $d$
ranging from 2 to 29, where the topmost line for $d=2$ also
corresponds to the optimal trade-off. For comparison, the
trade-off for the Gauss pointer (dashed) and square pointer
(dash-dotted) are also shown in this figure.} \label{fig:1}
% Give a unique label
\end{figure}

In Fig. \ref{fig:1} we give a plot of the trade-off between the
quality factor $F$ and precision $G$ for the unsharp measurements
of Eq. \eqref{eq3-1} with the first ten primes, and for comparison,
we also show the trade-off for the Gauss pointer considered
usually and the simple square pointer \cite{shareBT1}. For $d=2$,
as mentioned before, it saturates the optimal trade-off constraint
$F^2+G^2=1$, i.e., for any $F$, there exists optimal measurement
pointer that achieves the maximum $G$ \cite{shareBT1}. For the
prime $d\geqslant3$, as shown in Fig. \ref{fig:1}, although the
corresponding trade-off is not optimal, it is still better than
that given by the Gauss pointer for any $F$ (if $d=3$) or when
$F$ is smaller than a threshold (if $d\geqslant 5$), and with an
increase in $d$, it approaches gradually the trade-off $F+G=1$
given by the square pointer \cite{shareBT1}.

\section{Sharing NAQC by sequential observers} \label{sec:4}
%%%%%%%%%%%%%%%%%%%%%%%%%%%%%%%%%%%%%%%%%%%%%%%%%%%%%%%%%%%%%%%%%%
To address the question for sequential sharing of the NAQC, we
consider a scenario in which multiple Alices (say, Alice$_1$,
Alice$_2$, etc.) have access to half of an entangled qudit pair
and a spatially separated single Bob has access to the other half,
and they agree on the measurement settings $\{E^v\}$ in prior
\cite{{shareBT1}}. First, Alice$_1$ and Bob share the state
$\rho_{A_1B}$ and Alice$_1$ proceeds by choosing randomly one of
$\{E^v\}$ and performs the unsharp measurements on qudit $A_1$.
She then passes the measured qudit (we rename it as qudit $A_2$)
on to Alice$_2$ who measures again and passes it on to Alice$_3$,
and so on until the last Alice. During the whole process, every
Alice is assumed to be ignorant of the measurement settings chosen
by the former Alices, that is, communications among them are
forbidden and each Alice chooses independently and randomly one
of the measurement setting. Our aim is to determine the maximum
number of Alices whose statistics of measurements can demonstrate
NAQC with a spatially separated single Bob.

\subsection{Sharing NAQC between Alice$_1$ and Bob} \label{sec:4a}
%%%%%%%%%%%%%%%%%%%%%%%%%%%%%%%%%%%%%%%%%%%%%%%%%%%%%%%%%%%%%%%%%%
For the given $\rho_{A_1B}$, Alice$_1$ proceeds by choosing
randomly the measurement setting $E^v$, performing the unsharp
measurements \eqref{eq3-1} on $A_1$ and recording her outcomes.
Then within the L\"{u}ders rule \cite{Luders}, the selective
postmeasurement states can be written as
%Then the selective postmeasurement states can be described by the L\"{u}ders rule
%%%%%%%%%%%%%%%%%%%%%%%%%%%%%
\begin{equation}\label{eq3-3}
 \rho_{A_1 B|E^v_a}= \left(\sqrt{E^v_a} \otimes \iden\right) \rho_{A_1B}
                     \left(\sqrt{E^v_a} \otimes \iden\right) \big/ p_{B|E^v_a},
\end{equation}
%%%%%%%%%%%%%%%%%%%%%%%%%%%%%
where $p_{B|E^v_a}=\tr [(E^v_a \otimes\iden)\rho_{A_1B}]$ is the
probability of the measurement outcome $a$, and the square roots
of the unsharp measurements can be obtained as \cite{shareST2}
%%%%%%%%%%%%%%%%%%%%%%%%%%%%%
\begin{equation}\label{eq3-4}
\begin{aligned}
 &\sqrt{E^v_a}= \left( \sqrt{\frac{1+d_1\lambda}{d}}-\sqrt{\frac{1-\lambda}{d}}\right)\Pi^v_a
                +\sqrt{\frac{1-\lambda}{d}}\iden,
\end{aligned}
\end{equation}
%%%%%%%%%%%%%%%%%%%%%%%%%%%%%
where as said before, we have defined $d_1=d-1$.

%Note that the NAQC takes its maximum for the maximally entangled states. Hence, to proceed, we consider
%the case that Alice$_1$ and Bob initially share the following maximally entangled two-qudit state
To proceed, we suppose that Alice$_1$ and Bob initially share the
following maximally entangled two-qudit state
%%%%%%%%%%%%%%%%%%%%%%%%%%%%%
\begin{equation}\label{eq3a-1}
 |\Psi\rangle_{A_1B}= \frac{1}{\sqrt{d}}\sum_{k=0}^{d-1} |kk\rangle,
\end{equation}
%%%%%%%%%%%%%%%%%%%%%%%%%%%%%
then from Eq. \eqref{eq3-3} one can obtain the postmeasurement
states $\rho_{A_1B|E^v_a}$ contingent upon Alice$_1$'s unsharp
measurement $E^v$ on $A_1$ with outcome $a$. By further tracing
over $A_1$ one can obtain the conditional states of qudit $B$ as
%%%%%%%%%%%%%%%%%%%%%%%%%%%%%
\begin{equation}\label{eq3a-2}
\begin{aligned}
 &\rho_{B|E^0_a}= \frac{1-\lambda_1}{d}\iden
                  +\lambda_1 |a\rangle \langle a|, \\
 &\rho_{B|E^d_a}= \frac{1}{d}\iden
                  +\frac{\lambda_1}{d}
                  \sum_{n_{1,2}=0 \atop n_1\neq n_2}^{d-1}
                  e^{i\frac{2\pi}{d}a(n_2-n_1)}
                  |n_1\rangle \langle n_2|, \\
 &\rho_{B|E^r_a}= \frac{1}{d}\iden
                  +\frac{\lambda_1}{d}
                  \sum_{n_{1,2}=0 \atop n_1\neq n_2}^{d-1}
                  e^{i\frac{2\pi}{d}r\theta_{a,n_{1,2}}}
                  |n_1\rangle \langle n_2|,
\end{aligned}
\end{equation}
%%%%%%%%%%%%%%%%%%%%%%%%%%%%%
where $\lambda_1$ is the sharpness parameter of Alice$_1$,
$r=1,\ldots,d-1$, and the probability of obtaining
$\rho_{B|E^v_a}$ is given by $p_{B|E^v_a}=1/d$ ($\forall v,a$).
Besides, we have defined $\theta_{a,n_{1,2}}= (a+n_2)^2-
(a+n_1)^2$ for convenience of later presentation.

%%%%%%%%%%%%%%%%%%%%%%%%%%%
\begin{table}[!h]
\tabcolsep 0pt
\caption{Bob's reference basis to measure the coherence of the collapsed states on qudit $B$, where $r=1, \ldots, d-1$.} \label{tab:1}
\vspace{-12pt}
\begin{center}
\renewcommand\arraystretch{1.20}
\def\temptablewidth{0.48\textwidth}
{\rule{\temptablewidth}{1pt}}
\begin{tabular*}{\temptablewidth}{@{\extracolsep{\fill}}cclll}
  \rm{Bob's collapsed state} & \rm{Bob's reference basis} \\ \hline
  $\rho_{B|E^0_a}$           & $\{|\phi^d_a\rangle\}$ \\
  $\rho_{B|E^d_a}$           & $\{|\phi^0_a\rangle\}$ \\
  $\rho_{B|E^r_a}$           & $\{|\phi^r_a\rangle\}$
\end{tabular*}
{\rule{\temptablewidth}{1pt}}
\end{center}
\end{table}
%%%%%%%%%%%%%%%%%%%%%%%%%%%

The states $\rho_{B|E^0_a}$, $\rho_{B|E^d_a}$, and
$\rho_{B|E^r_a}$ ($r=1, \ldots, d-1$) are diagonal in
the bases $\{|\phi^0_a\rangle\}$, $\{|\phi^d_a\rangle\}$, and
$\{|\phi^{d-r}_a\rangle\}$, respectively. For any $\rho_{B|E^v_a}$,
its coherence in the basis $\{|\phi^u_a\rangle\}$ with
$u\in\{0,1,\ldots,d\}$ are the same apart from the one mentioned
above under which it is diagonal. So for convenience of later
calculations, we assume that Bob chooses the bases
$\{|\phi^d_a\rangle\}$, $\{|\phi^0_a\rangle\}$, and
$\{|\phi^r_a\rangle\}$ to measure the coherences of
$\rho_{B|E^0_a}$, $\rho_{B|E^d_a}$, and $\rho_{B|E^r_a}$,
respectively (see Table \ref{tab:1}). Then by transforming the
collapsed states given in Eq. \eqref{eq3a-2} to the bases shown
in Table \ref{tab:1}, one can obtain
%%%%%%%%%%%%%%%%%%%%%%%%%%%%%
\begin{equation}\label{eq3a-3}
\begin{aligned}
 &\varrho_{B|E^0_a}= \frac{1}{d}\iden
                  +\frac{\lambda_1}{d}
                  \sum_{n_{1,2}=0 \atop n_1\neq n_2}^{d-1}
                  e^{i\frac{2\pi}{d}a(n_2-n_1)}
                  \left|\phi_{n_1}^d\right\rangle \left\langle\phi_{n_2}^d\right|, \\
 &\varrho_{B|E^d_a}= \frac{1}{d}\iden
                  +\frac{\lambda_1}{d}
                  \sum_{n_{1,2}=0 \atop n_1\neq n_2}^{d-1}
                  e^{i\frac{2\pi}{d}a(n_2-n_1)}
                  \left|\phi_{n_1}^0\right\rangle \left\langle\phi_{n_2}^0\right|, \\
 &\varrho_{B|E^r_a}= \frac{1}{d}\iden
                  +\frac{\lambda_1}{d^2}
                  \sum_{n_{1,2}=0 \atop n_1\neq n_2}^{d-1}
                  \sum_{k_{1,2}=0}^{d-1}
                  e^{i\frac{2\pi}{d}r\varphi_{a,k_{1,2},n_{1,2}}}
                  \left|\phi_{n_1}^r\right\rangle \left\langle\phi_{n_2}^r\right|,
\end{aligned}
\end{equation}
%%%%%%%%%%%%%%%%%%%%%%%%%%%%%
where $\varphi_{a,k_{1,2},n_{1,2}}=\theta_{a,k_{1,2}}+(k_2+n_2)^2
-(k_1+n_1)^2$.

The eigenvalues of $\rho_{B|E^v_a}$ ($\forall v,a$) are
$\epsilon_0= (1+d_1\lambda_1)/d$ with degeneracy $1$ and
$\epsilon_1= (1-\lambda_1)/d$ with degeneracy $d_1$. Then
in the scenario where Alice$_1$ chooses $E^v$ with probability
$p_{A|E^v}=1/(d+1)$ ($\forall v$), the two forms of ASC
attainable by Bob could be obtained as
%%%%%%%%%%%%%%%%%%%%%%%%%%%%%
\begin{equation}\label{eq3a-4}
\begin{aligned}
  \mathcal{N}^{l_1}_{A_1 B}= \, & (d^2-1)\lambda_1, \\
  \mathcal{N}^{re}_{A_1 B}= \, & \frac{d+1}{d}\big[(1+d_1\lambda_1)\log_2(1+d_1 \lambda_1) \\
                             & +d_1(1-\lambda_1)\log_2(1-\lambda_1)\big].
\end{aligned}
\end{equation}
%%%%%%%%%%%%%%%%%%%%%%%%%%%%%

\begin{figure}
\centering
\resizebox{0.44 \textwidth}{!}{%
\includegraphics{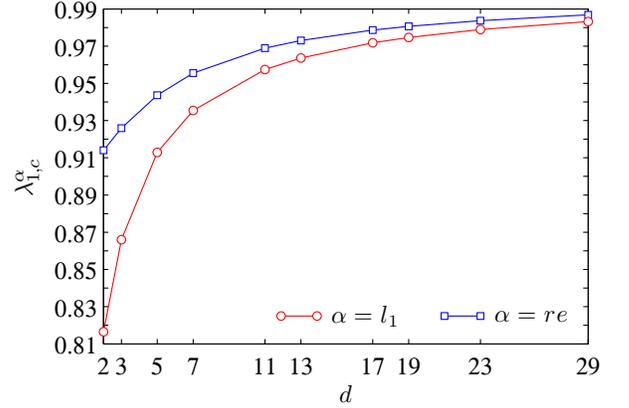}}
% If not, use\vspace{5cm}
% Give the correct figure height in cm
\caption{The critical sharpness parameter $\lambda_{1,c}^\alpha$
($\alpha=l_1$ or $re$) stronger than which Alice$_1$ can demonstrate
NAQC with Bob versus the prime $d$.} \label{fig:2}
% Give a unique label
\end{figure}

From Eq. \eqref{eq3a-4} one can obtain the critical sharpness
parameter $\lambda_{1,c}^{l_1}= \sqrt{d/(d+1)}$ stronger than
which Alice$_1$ can demonstrate the $l_1$ norm of NAQC with Bob.
Similarly, one can obtain numerically the critical
$\lambda_{1,c}^{re}$ stronger than which Alice$_1$ can
demonstrate the relative entropy of NAQC with Bob. In Fig.
\ref{fig:2}, we give a plot of the critical $\lambda_{1,c}^\alpha$
versus the prime $d$. It is clear that it always increases with
an increase in $d$. Besides, it follows from Eqs. \eqref{eq3-n4}
and \eqref{eq3a-4} that for any fixed prime $d$,
$\mathcal{N}^{\alpha}_{A_1 B}$ ($\alpha=l_1$ or $re$) is solely
determined by the precision of the unsharp measurements of
Alice$_1$.

\subsection{Sharing NAQC between Alice$_2$ and Bob} \label{sec:4b}
%%%%%%%%%%%%%%%%%%%%%%%%%%%%%%%%%%%%%%%%%%%%%%%%%%%%%%%%%%%%%%%%%%
To proceed, we see whether the measurement statistics of two
Alices can demonstrate NAQC with Bob. As we consider a sequential
steering scenario, after finishing the unsharp measurement $E^v$,
Alice$_1$ passes the measured qudit $A_1$ on to Alice$_2$ who is
independent of her, namely, the classical information regarding
the measurement setting and the outcome of Alice$_1$ is not
conveyed. Then according to the L\"{u}ders transformation rule
\cite{Luders}, the state she shared with Bob can be written as
%%%%%%%%%%%%%%%%%%%%%%%%%%%%%
\begin{equation}\label{eq3b-1}
 \rho_{A_2B|E^v}= \sum_{a=0}^{d-1} \left(\sqrt{E^v_a}\otimes\iden\right)
                  \rho_{A_1B} \left(\sqrt{E^v_a}\otimes\iden \right),
\end{equation}
%%%%%%%%%%%%%%%%%%%%%%%%%%%%%
where we have denoted by $\rho_{A_2B|E^v}$ the output state of
Alice$_1$'s unsharp measurements $E^v=\{E^v_a\}$. Then for
$\rho_{A_1B}$ of Eq. \eqref{eq3a-1}, one can obtain
%%%%%%%%%%%%%%%%%%%%%%%%%%%%%
\begin{equation}\label{eq3b-2}
\begin{aligned}
 &\rho_{A_2B|E^0}= \lambda_0 \rho_{A_1B}
                   +\frac{1-\lambda_0}{d}
                   \sum_{n=0}^{d-1}
                   |nn\rangle \langle nn|, \\
 &\rho_{A_2B|E^d}= \lambda_0 \rho_{A_1B}
                   + \frac{1-\lambda_0}{d^3}
                   \sum_{n,n_{1,2}, \atop k_{1,2}=0}^{d-1}
                   e^{i\frac{2\pi}{d}\xi_{n,n_{1,2},k_{1,2}}}
                   |n_1n_2\rangle \langle k_1k_2|, \\
 &\rho_{A_2B|E^r}= \lambda_0 \rho_{A_1B}
                   + \frac{1-\lambda_0}{d^3}
                   \sum_{n,n_{1,2}, \atop k_{1,2}=0}^{d-1}
                   e^{i\frac{2\pi}{d}r\zeta_{n,n_{1,2},k_{1,2}}}
                   |n_1n_2\rangle \langle k_1k_2|, \\
\end{aligned}
\end{equation}
%%%%%%%%%%%%%%%%%%%%%%%%%%%%%
where $\lambda_0$ can be obtained directly by substituting
$\lambda$ in Eq. \eqref{eq3-n3} with $\lambda_1$ and we have
defined
%%%%%%%%%%%%%%%%%%%%%%%%%%%%%
\begin{equation}\label{eq3b-3}
\begin{aligned}
 & \xi_{n,n_{1,2},k_{1,2}}=n(n_1-n_2-k_1+k_2), \\
 & \zeta_{n,n_{1,2},k_{1,2}}= \theta_{n,n_{1,2}}-\theta_{n,k_{1,2}}.
\end{aligned}
\end{equation}
%%%%%%%%%%%%%%%%%%%%%%%%%%%%%

After receiving the qudit from Alice$_1$, Alice$_2$ performs
the measurements $\{E^v_a\}$ on it (we rename it as qudit
$A_2$) with the sharpness parameter $\lambda_2$. As Alice$_2$ is
assumed to be ignorant of the measurement setting chosen by
Alice$_1$ when measuring the qudit which is now in her possession,
she has to consider the average effect of all possible measurement
settings of Alice$_1$, that is, the NAQC Alice$_2$ shared with Bob
has to be averaged over the $d+1$ possible outputs of Alice$_1$
given in Eq. \eqref{eq3b-2}.

First, for $\rho_{A_2B|E^0}$, one can obtain the selective
postmeasurement states $\rho_{B|E^0E^v_a}$ ($v=0,1,\ldots,d$) of Bob after
Alice$_2$'s unsharp measurement $ \{E^v_a\}$ on qudit $A_2$,
whose forms are similar to $\rho_{B|E^v_a}$  in Eq. \eqref{eq3a-2}.
To be explicit, one could obtain $\rho_{B|E^0E^0_a}$
($\rho_{B|E^0E^d_a}$ and $\rho_{B|E^0E^r_a}$) by substituting the
parameter $\lambda_1$ in $\rho_{B|E^0_a}$ ($\rho_{B|E^d_a}$ and
$\rho_{B|E^r_a}$) with $\lambda_2$ ($\lambda_0\lambda_2$). Thereby, the
$l_1$ norm of coherence for $\rho_{B|E^0E^0_a}$ is $d_1\lambda_2$
and that for both $\rho_{B|E^0E^d_a}$ and $\rho_{B|E^0E^r_a}$ is
$d_1\lambda_0\lambda_2$. Thus the $l_1$ norm of ASC attainable
from $\rho_{A_2B|E^0}$ can be obtained as
%%%%%%%%%%%%%%%%%%%%%%%%%%%%%
\begin{equation}\label{eq3b-4}
 \mathcal{N}^{l_1}_{A_2 B|E^0}(\rho_{A_2B|E^0})= d_1(1+d\lambda_0)\lambda_2.
\end{equation}
%%%%%%%%%%%%%%%%%%%%%%%%%%%%%
Moreover, the eigenvalues of $\rho_{B|E^0 E^0_a}$ can be obtained
as $\varepsilon_0= (1+d_1\lambda_2)/d$ with degeneracy $1$ and
$\varepsilon_1= (1-\lambda_2)/d$ with degeneracy $d_1$, while
those for $\rho_{B|E^0 E^d_a}$ and $\rho_{B|E^0 E^r_a}$ can be
obtained directly by substituting $\lambda_2$ in
$\varepsilon_{0,1}$ with $\lambda_0\lambda_2$, hence the relative
entropy of ASC for $\rho_{A_2B|E^0}$ can be obtained as
%%%%%%%%%%%%%%%%%%%%%%%%%%%%%
\begin{equation}\label{eq3b-5}
\begin{aligned}
 \mathcal{N}^{re}_{A_2 B|E^0}(\rho_{A_2B|E^0})= \, & (1+d)\log_2 d -H_2\left(\frac{1+d_1\lambda_2}{d}\right) \\
                                              & -d_1 \left(1-\lambda_0\lambda_2+\frac{1-\lambda_2}{d}\right)\log_2 d_1 \\
                                              & -dH_2\left(\frac{1+d_1\lambda_0\lambda_2}{d}\right),
\end{aligned}
\end{equation}
%%%%%%%%%%%%%%%%%%%%%%%%%%%%%
where $H_2(\cdot)$ is the binary Shannon entropy function.

Next, for $\rho_{A_2B|E^d}$ of Eq. \eqref{eq3b-2}, the selective
postmeasurement state $\rho_{B|E^d E^d_a}$ of qudit $B$ after
Alice$_2$'s measurement $\{E^d_a\}$ is similar to
$\rho_{B|E^d_a}$ in Eq. \eqref{eq3a-2}, with however the parameter
$\lambda_1$ being replaced by $\lambda_2$, while the
postmeasurement state $\rho_{B|E^d E^0_a}$ ($\rho_{B|E^d E^r_a}$)
of Bob after Alice$_2$'s measurement $\{E^0_a\}$ ($\{E^r_a\}$) is
similar to $\rho_{B|E^0_a}$ ($\rho_{B|E^r_a}$) of Eq.
\eqref{eq3a-2}, with however the parameter $\lambda_1$ being
replaced by $\lambda_0\lambda_2$. As a result, the $l_1$ norm and
relative entropy of ASCs attainable from $\rho_{A_2B|E^d}$ have the
same form as that given in Eqs. \eqref{eq3b-4} and \eqref{eq3b-5},
respectively.

Finally, for $\rho_{A_2B|E^r}$ with $r=1,\ldots,d-1$, the
selective postmeasurement state $\rho_{B|E^r E^0_a}$
($\rho_{B|E^r E^d_a}$) of qudit $B$ after Alice$_2$'s measurement
$\{E^0_a\}$ ($\{E^d_a\}$) on qudit $A_2$ is similar to
$\rho_{B|E^0_a}$ ($\rho_{B|E^d_a}$) in Eq. \eqref{eq3a-2},
with however the parameter $\lambda_1$ being replaced by
$\lambda_0\lambda_2$. In addition, the selective postmeasurement
state $\rho_{B|E^r E^s_a}$ ($s=1,\ldots,d-1$) of qudit $B$ after
Alice$_2$'s measurement $\{E^s_a\}$ on qudit $A_2$ can be obtained
as
%%%%%%%%%%%%%%%%%%%%%%%%%%%%%
\begin{equation}\label{eq3b-6}
\begin{aligned}
\rho_{B|E^r E^s_a}=\, & \frac{1}{d}\iden
                        +\frac{\lambda_2}{d}
                        \sum_{n_{1,2}=0 \atop n_1\neq n_2}^{d-1}
                        \Bigg[\lambda_0
                        e^{i\frac{2\pi}{d}s\theta_{a,n_1,n_2}} \\
                      & +\frac{1-\lambda_0}{d^2}\sum_{n,k_{1,2}=0}^{d-1}
                        e^{i\frac{2\pi}{d}\varsigma_{a,n,k_{1,2},n_{1,2}}} \Bigg]
                        |n_1\rangle\langle n_2|,
\end{aligned}
\end{equation}
%%%%%%%%%%%%%%%%%%%%%%%%%%%%%
then by transforming it to the reference basis
$\{|\phi^s_a\rangle\}$ (see Table \ref{tab:1}), one has
%%%%%%%%%%%%%%%%%%%%%%%%%%%%%
\begin{equation}\label{eq3b-7}
\begin{aligned}
 \varrho_{B|E^r E^s_a}= \, & \frac{1}{d}\iden
                               +\frac{\lambda_2}{d^2}
                               \sum_{n_{1,2}=0 \atop n_1\neq n_2}^{d-1}
                               \Bigg[\lambda_0
                               \sum_{k_{1,2}=0}^{d-1}
                               e^{i\frac{2\pi}{d}s\varphi_{a,k_{1,2},n_{1,2}}} \\
                             & +\frac{1-\lambda_0}{d^2}
                               \sum_{n,j_{1,2} \atop k_{1,2}=0}^{d-1}
                               e^{i\frac{2\pi}{d}\chi_{a,n,j_{1,2},k_{1,2},n_{1,2}}} \Bigg]
                               \left|\phi_{n_1}^s\right\rangle \left\langle \phi_{n_2}^s\right|,
\end{aligned}
\end{equation}
%%%%%%%%%%%%%%%%%%%%%%%%%%%%%
where we have defined
%%%%%%%%%%%%%%%%%%%%%%%%%%%%%
\begin{equation}\label{eq3b-8}
\begin{aligned}
 \varsigma_{a,n,k_{1,2},n_{1,2}}= \, & s\theta_{a,k_{1,2}}- r(\theta_{n,k_{1,2}}+\theta_{n,n_{1,2}}), \\
 \chi_{a,n,j_{1,2},k_{1,2},n_{1,2}}= \, & r(\theta_{n,k_{1,2}}-\theta_{n,j_{1,2}})+s[(a+k_2)^2 \\
                                        & -(a+j_2)^2+(n_2+k_1)^2 - (n_1+j_1)^2].
\end{aligned}
\end{equation}
%%%%%%%%%%%%%%%%%%%%%%%%%%%%%

For $\varrho_{B|E^r E^s_a}$ of Eq. \eqref{eq3b-7}, the $l_1$ norm
of coherence is given by $d_1\lambda_2$ for $s= r$ and
$d_1\lambda_0\lambda_2$ for $s\neq r$. Similarly, the eigenvalues
of $\varrho_{B|E^r E^s_a}$ for $s= r$ are $\varepsilon_0=
(1+d_1\lambda_2)/d$ with degeneracy $1$ and $\varepsilon_1=
(1-\lambda_2)/d$ with degeneracy $d-1$, while the eigenvalues of
$\varrho_{B|E^r E^s_a}$ for $s\neq r$ could be obtained directly
by substituting $\lambda_2$ in $\varepsilon_{0,1}$ with
$\lambda_0\lambda_2$. Then after some algebra, one can obtain that
the $l_1$ norm and relative entropy of ASCs attainable from
$\rho_{A_2B|E^r}$ also have the same form as that given in Eqs.
\eqref{eq3b-4} and \eqref{eq3b-5}, respectively.

Since we are concerned with an unbiased input scenario, all the
possible measurement settings of Alice$_1$ are equiprobable, i.e.,
her probability of choosing the measurement setting $E^v$ is
$p_{A|E^v}= 1/(d+1)$ ($\forall v$), thus the steerable coherence
for Alice$_2$ and Bob can be obtained as
%%%%%%%%%%%%%%%%%%%%%%%%%%%%%
\begin{equation}\label{eq3b-9}
 \mathcal{N}^{\alpha}_{A_2 B}= \sum_v p_{A|E^v} \mathcal{N}^{\alpha}_{A_2 B|E^v}(\rho_{A_2B|E^v}),
\end{equation}
%%%%%%%%%%%%%%%%%%%%%%%%%%%%%
where $\alpha=l_1$ or $re$. By substituting $p_{A|E^v}= 1/(d+1)$
and the associated ASC for $\rho_{A_2B|E^0}$,
$\rho_{A_2B|E^d}$, and $\rho_{A_2B|E^r}$ ($r=1,\ldots,d-1$) into
Eq. \eqref{eq3b-9}, one can obtain
%%%%%%%%%%%%%%%%%%%%%%%%%%%%%
\begin{equation}\label{eq3b-10}
\begin{aligned}
 \mathcal{N}^{l_1}_{A_2 B}=\,& d_1(1+d\lambda_0)\lambda_2, \\
 \mathcal{N}^{re}_{A_2 B}=\, & (1+d)\log_2 d- d_1 \left(1-\lambda_0\lambda_2+\frac{1-\lambda_2}{d}\right)\log_2 d_1 \\
                             & -H_2\left(\frac{1+d_1\lambda_2}{d}\right)
                               -dH_2\left(\frac{1+d_1\lambda_0\lambda_2}{d}\right).
\end{aligned}
\end{equation}
%%%%%%%%%%%%%%%%%%%%%%%%%%%%%

\begin{figure}
\centering
\resizebox{0.44 \textwidth}{!}{%
\includegraphics{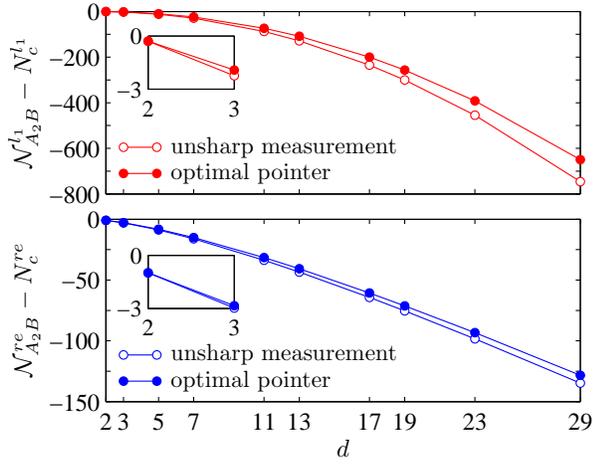}}
% If not, use\vspace{5cm}
% Give the correct figure height in cm
\caption{$\mathcal{N}^{\alpha}_{A_2 B}-N^\alpha_{c}$ ($\alpha=l_1$
or $re$) versus the prime $d$ with $\lambda_1=\lambda_{1,c}^\alpha$
and $\lambda_2=1$. Their dependence on small $d$ are also shown in
the insets to better visual the behavior.} \label{fig:3}
% Give a unique label
\end{figure}

As mentioned before, $\lambda_0$ depends on $\lambda_1$, so
$\mathcal{N}^{\alpha}_{A_2 B}$ ($\alpha=l_1$ or $re$) for
Alice$_2$ and Bob is determined by both the quality factor of
Alice$_1$'s measurements and the precision of Alice$_2$'s
measurements. Eq. \eqref{eq3b-10} also reveals that
$\mathcal{N}^{\alpha}_{A_2 B}$ decreases with the increasing
disturbance (i.e., increasing precision) of Alice$_1$'s
measurements and for any fixed disturbance of Alice$_1$, Alice$_2$
can enhance $\mathcal{N}^{\alpha}_{A_2 B}$ by improving the
precision of her measurements. Under the condition of guaranteeing
the observation of NAQC for Alice$_1$ and Bob,
$\mathcal{N}^{\alpha}_{A_2 B}$ takes its maximum when the
measurement of Alice$_2$ is sharp (i.e., $\lambda_2=1$) and the
sharpness parameter $\lambda_1$ of Alice$_1$ is a slightly
stronger than $\lambda_{1,c}^\alpha$. In Fig. \ref{fig:3} we show
the $d$ dependence of $\mathcal{N}^{\alpha}_{A_2 B}-N^\alpha_{c}$
with $\lambda_1=\lambda_{1,c}^\alpha$ and $\lambda_2=1$ (the
hollow circles). It can be seen that it decreases with the
increase of the prime $d$ and is always smaller than 0. This
indicates that when Alice$_1$ steers successfully the NAQC on
Bob's side, Alice$_2$ will cannot steer it again.

\begin{figure}
\centering
\resizebox{0.44 \textwidth}{!}{%
\includegraphics{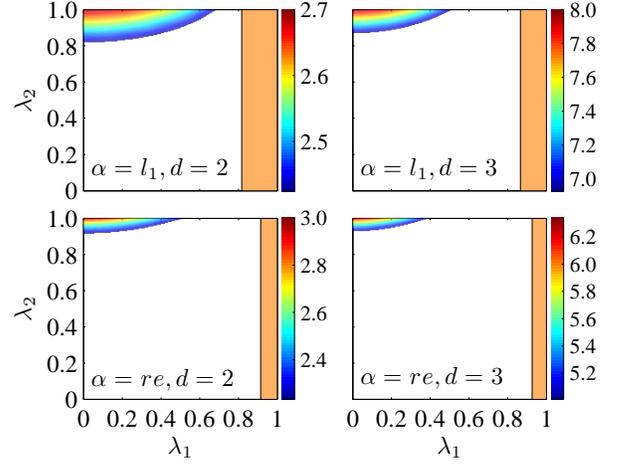}}
% If not, use\vspace{5cm}
% Give the correct figure height in cm
\caption{Plot of $\mathcal{N}^{\alpha}_{A_2 B}$ ($\alpha=l_1$ or $re$)
in the parameter region $(\lambda_1,\lambda_2)$ in which Alice$_2$
can steer the NAQC on Bob for $d=2$ and 3, and in the orange shaded
regions, Alice$_1$ can steer the NAQC on Bob (note that Alice$_1$'s
ability to steer the NAQC is independent of $\lambda_2$).} \label{fig:4}
% Give a unique label
\end{figure}

Having clarified the fact that Alice$_2$ cannot steer the NAQC
on qudit $B$ when Alice$_1$ steers it successfully, it is definite
that  the subsequent Alices (i.e., Alice$_n$ with $n\geqslant 3$)
also can never steer the NAQC. Now, the issue that remains is
whether Alice$_2$ can steer the NAQC after Alice$_1$'s unsharp
measurements with $0<\lambda_1\leqslant\lambda_{1,c}^\alpha$, namely,
Alice$_1$'s measurements are unable to extract enough information
to observe the NAQC. To address this precisely, we further display
in Fig. \ref{fig:4} dependence of $\mathcal{N}^{\alpha}_{A_2 B}$
($\alpha=l_1$ or $re$) on $\lambda_1$ and $\lambda_2$ for two
primes $d=2$ and 3, and in the same figure we also show the regions
of $\lambda_1$ in which Alice$_1$ can steer the NAQC on qudit $B$.
Looking at this figure, one can see that there exists parameter
region $(\lambda_1,\lambda_2)$ in which Alice$_2$ can steer the
NAQC, and to ensure Alice$_2$'s steerability of the NAQC, the
sharpness parameter $\lambda_1$ of Alice$_1$'s measurements should
be weaker than a threshold $\lambda_{1,t}^\alpha$. From Fig.
\ref{fig:4} one can also note that such a parameter region shrinks
with an increase in the prime $d$. In particular, in the region
of $\lambda_1\in [\lambda_{1,t}^\alpha,\lambda_{1,c}^\alpha]$,
both Alice$_1$ and Alice$_2$ cannot steer the NAQC, that is,
Alice$_1$ should tune the sharpness of her measurement,
as an unappropriate strength of measurement will prevent both of
them from demonstrating the NAQC. All these show evidently that
no matter how the sharpness parameters $\lambda_1$ and $\lambda_2$
are chosen, at most one Alice (i.e., Alice$_1$ or
Alice$_2$) can demonstrate NAQC with a spatially separated Bob.

By comparing Eqs. \eqref{eq3a-4} and \eqref{eq3b-10} one can
further note that $\mathcal{N}^{\alpha}_{A_1 B}$ and
$\mathcal{N}^{\alpha}_{A_2 B}$ show an opposite dependence on
$\lambda_1$. That is to say, the enhancement of Alice$_1$'s
steerable coherence implies the degradation of Alice$_2$'s, and
vice versa. In some sense, one may recognize this as a kind of
sequentially monogamous characteristics of NAQC, as it sets limit
on the possibility for Alice$_1$ and Alice$_2$ to steer the
NAQC with Bob simultaneously, even if the resource state is
maximally entangled.

Note that for the $d=2$ case, our results also apply to NAQC
captured by the criterion of Eq. \eqref{eq2-3}, the sequential
sharing of which has been discussed in Ref. \cite{sharenaqc,
sharenaqce}. But there are minor errors in \cite{sharenaqce}
related to the relative entropy of NAQC.

As we showed above, for the unbiased inputs of Alice$_2$ (i.e.,
equiprobable measurement settings for Alice$_1$), it is impossible
for the NAQC of an entangled pair of qudits be distributed between
two Alices who act sequentially and independently of each other.
Then another interesting question to ask is whether Alice$_2$
could demonstrate NAQC when the inputs to her are biased. This
question is important by itself as Alice$_2$ is ignorant of
Alice$_1$'s measurement setting, thereby her premeasurement state
is a mixture of the collapsed states of Alice$_1$'s possible
measurements weighted by their probabilities. To answer this
question, we resort again to the amounts of steered coherence
discussed above for $\rho_{A_2B|E^v}$ ($v=0,1,\ldots,d$) of Eq.
\eqref{eq3b-2}. As mentioned before, the different
$\rho_{A_2B|E^v}$ yields the same steerable coherence, thus even
there is input bias for Alice$_2$, it does not change the limit
that it is impossible for more than one Alice to demonstrate
NAQC with Bob.

It is also relevant to ask whether the above conclusion also holds
in a scenario where Alice$_1$ measures the qudit $A_1$ with unequal
sharpness, e.g., Alice$_1$ performs the measurement $E^v$ with
sharpness parameter $\lambda_{1,v}$. In this case, following the
similar derivations as in the previous sections, one can show that
Alice$_2$ still cannot demonstrate NAQC with Bob if Alice$_1$ can
do so. For example, when $d=2$ and Alice$_2$'s measurement is
sharp, one has
%%%%%%%%%%%%%%%%%%%%%%%%%%%%%
\begin{equation}\label{eq3b-11}
\begin{aligned}
 \mathcal{N}_{A_1B}^{l_1}= \sum_v \lambda_{1,v},~
 \mathcal{N}_{A_2B}^{l_1}= 1+\frac{2}{3}\sum_v \sqrt{1-\lambda_{1,v}^2},
\end{aligned}
\end{equation}
%%%%%%%%%%%%%%%%%%%%%%%%%%%%%
then one can show that the maximal NAQC for Alice$_2$ and Bob corresponds
to $\lambda_{1,v} =\lambda_1$ ($\forall v$). Thus she still cannot
demonstrate NAQC with Bob.

One may also be concerned with the issue that whether the
number of Alices sequentially sharing the NAQC could be enhanced
when we consider the weak measurement with optimal pointer
\cite{shareBT1}. For $d=2$, as said, it is already optimal. For
the general prime $d$, by using Eq. \eqref{eq3-m3} and after some
algebra similar to those for the unsharp measurements, it can be
found that  although $\mathcal{N}^{\alpha}_{A_2 B}$ can be enhanced
to some extent (see the solid circles in Fig. \ref{fig:3}), it
still cannot exceed $N^{\alpha}_c$ under the condition of
$\mathcal{N}^{\alpha}_{A_1 B}>N_c^\alpha$. This confirms again
that at most one Alice can demonstrate NAQC with Bob. But it
should be note that although there is no NAQC in the
postmeasurement states of Eq. \eqref{eq3b-2}, there are still
other forms of residual quantum correlations. For example, for
the $d=2$ case, when $\mathcal{N}_{A_1B}^{l_1}=2.50$, one can get
a $14.26\%$ violation of the CHSH inequality for all the
postmeasurement states of Alice$_1$ \cite{CHSH}.

Lastly, when $d$ is a power of a prime, a complete set of
$d+1$ mutually unbiased bases also exists \cite{MUB2}, and one can
show in a similar way that all the above results also apply to
this case. But as the NAQC were defined only for $d$ being a prime
or a prime power \cite{naqc2}, the formulation of NAQC and its
sequential sharing for a general $d$ are still open questions.

\section{Conclusion} \label{sec:5}
%%%%%%%%%%%%%%%%%%%%%%%%%%%%%%%%%%%%%%%%%%%%%%%%%%%%%%%%%%%%%%%%%%
In conclusion, we have investigated sequential sharing of NAQC in
the $(d\times d)$-dimensional (i.e., two-qudit) state, with $d$
being a power of a prime. We consider these high-dimensional
states, as compared with the two-dimensional ones, not only
enrich our comprehension of the nonlocal characteristics in quantum
theory but also show many
advantages in quantum communication tasks such as the high
channel capacity and security \cite{adv1,adv2,adv3,adv4,adv5}. By
considering a scenario in which multiple Alices perform their
unsharp measurements sequentially and independently of each other
on the same half of an entangled qudit pair and a single Bob
measures coherence of the collapsed states on the other half,
we showed that for both the metrics (i.e., the $l_1$ norm and
relative entropy) used for quantifying coherence and for both the
unbiased and biased input scenarios, at most one Alice can
demonstrate NAQC with Bob. Moreover, we showed that the conclusion
also holds even when one considers the weak measurements with the optimal
pointer, even when Alice$_1$'s measurement settings are biased, or
when she measures the qudit with unequal sharpness associated with
different measurement settings.

The results presented above indicate that there exists a strict
limit on the number of Alices whose statistics of measurements can
demonstrate NAQC with a spatially separated Bob. This provides an
alternative dimension in the context of sequential sharing of
quantum correlations and might shed light on the interplay between
quantum measurement and quantum correlations for high-dimensional
states. Furthermore, in the sense that the maximum number of
observers being able to sequentially sharing quantum correlations
is inherently related to the hierarchy of the strengths of quantum
correlations, for example, Bell-CHSH nonlocality could be shared
by not more than two unbiased observers
\cite{shareBT1,shareBT2,shareBT3} and EPR steering could be shared
by at most $n$ observers when the steering inequality based on $n$
measurement settings is used \cite{shareST1,shareST2}, it is
intuitive to conjecture that the observation that the NAQC can be
shared by at most one observer might indicate that it
characterizes a kind of quantum correlation which is stronger
than Bell nonlocality for the general two-qudit states, just as
that for the two-qubit states \cite{naqc3}. Of course, further
study is still needed to provide a rigorous proof of this
conjecture. As there are other coherence measures \cite{Plenio,Hu},
deriving the associated criteria for capturing NAQC and exploring
whether they could provide an advantage over those considered in
this work in the context of NAQC sharing is another direction for
future studies. Moreover, how such a strong quantum correlation
can be used in practical communication and computation tasks would
also be worth pursuing in the future.

\section*{ACKNOWLEDGMENTS}
This work was supported by the National Natural Science Foundation
of China (Grant Nos. 11675129 and 11934018), the Strategic Priority
Research Program of Chinese Academy of Sciences (Grant No. XDB28000000),
and Beijing Natural Science Foundation (Grant No. Z200009).

\newcommand{\PRL}{Phys. Rev. Lett. }
\newcommand{\RMP}{Rev. Mod. Phys. }
\newcommand{\PRA}{Phys. Rev. A }
\newcommand{\PRB}{Phys. Rev. B }
\newcommand{\PRD}{Phys. Rev. D }
\newcommand{\PRE}{Phys. Rev. E }
\newcommand{\PRX}{Phys. Rev. X }
\newcommand{\NJP}{New J. Phys. }
\newcommand{\JPA}{J. Phys. A }
\newcommand{\JPB}{J. Phys. B }
\newcommand{\OC}{Opt. Commun.}
\newcommand{\PLA}{Phys. Lett. A }
\newcommand{\EPJB}{Eur. Phys. J. B }
\newcommand{\EPJD}{Eur. Phys. J. D }
\newcommand{\NP}{Nat. Phys. }
\newcommand{\NC}{Nat. Commun. }
\newcommand{\EPL}{Europhys. Lett. }
\newcommand{\AdP}{Ann. Phys. (Berlin) }
\newcommand{\AoP}{Ann. Phys. (N.Y.) }
\newcommand{\QIC}{Quantum Inf. Comput. }
\newcommand{\QIP}{Quantum Inf. Process. }
\newcommand{\CPB}{Chin. Phys. B }
\newcommand{\IJTP}{Int. J. Theor. Phys. }
\newcommand{\IJQI}{Int. J. Quantum Inf. }
\newcommand{\IJMPB}{Int. J. Mod. Phys. B }
\newcommand{\PR}{Phys. Rep. }
\newcommand{\SR}{Sci. Rep. }
\newcommand{\LPL}{Laser Phys. Lett. }
\newcommand{\SCG}{Sci. China Ser. G }
\newcommand{\JMP}{J. Math. Phys. }
\newcommand{\RPP}{Rep. Prog. Phys. }
\newcommand{\PA}{Physica A }
\newcommand{\SCPMA}{Sci. China-Phys. Mech. Astron. }

%BibTeX users please use
%\bibliographystyle{}
%\bibliography{}

\begin{thebibliography}{50}
%Format for Journal Reference
% 01-10
\bibitem{Nielsen} M. A. Nielsen, and I. L. Chuang, \textit{Quantum Computation and Quantum Information} (Cambridge University Press, Cambridge, UK, 2010).
\bibitem{Bell1} Genovese M, \PR \textbf{413}, 319 (2005).
\bibitem{Bell2} N. Brunner, D. Cavalcanti, S. Pironio, V. Scarani, and S. Wehner, \RMP \textbf{86}, 419 (2014).
\bibitem{steer1} D. Cavalcanti, and P. Skrzypczyk, \RPP \textbf{80}, 024001 (2017).
\bibitem{steer2} R. Uola, A. C. S. Costa, H. C. Nguyen, and O. G\"{u}hne, \RMP \textbf{92}, 015001 (2020).
\bibitem{QE} R. Horodecki, P. Horodecki, M. Horodecki, and K. Horodecki, \RMP \textbf{81}, 865 (2009).
\bibitem{QD} K. Modi, A. Brodutch, H. Cable, Z. Paterek, and V. Vedral, \RMP \textbf{84}, 1655 (2012).
\bibitem{monoe} V. Coffman, J. Kundu, and W. K. Wootters, \PRA \textbf{61}, 052306 (2000).
\bibitem{monon} B. Toner, Proc. R. Soc. London A \textbf{465}, 59 (2009).
\bibitem{monos} M. D. Reid, \PRA \textbf{88}, 062108 (2013).

% 11-20
\bibitem{monoc} L. Lami, C. Hirche, G. Adesso, and A. Winter, \PRL \textbf{117}, 220502 (2016).
\bibitem{monod} A. Streltsov, G. Adesso, M. Piani, and D. Bru{\ss}, \PRL \textbf{109}, 050503 (2012).
\bibitem{shareBT1} R. Silva, N. Gisin, Y. Guryanova, and S. Popescu, \PRL \textbf{114}, 250401 (2015).
\bibitem{shareBT2} S. Mal, A. S. Majumdar, and D. Home, Mathematics \textbf{4}, 48 (2016).
\bibitem{shareBT3} C. Ren, T. Feng, D. Yao, H. Shi, J. Chen, and X. Zhou, \PRA \textbf{100}, 052121 (2019).
\bibitem{shareBT4} D. Das, A. Ghosal, S. Sasmal, S. Mal, and A. S. Majumdar, \PRA \textbf{99}, 022305 (2019).
\bibitem{shareBE1} M. J. Hu, Z. Y. Zhou, X. M. Hu, C. F. Li, G. C. Guo, and Y. S. Zhang, npj Quantum Inf. \textbf{4}, 63 (2018).
\bibitem{shareBE2} M. Schiavon, L. Calderaro, M. Pittaluga, G. Vallone, and P. Villoresi, Quantum Sci. Technol. \textbf{2}, 015010 (2017).
\bibitem{shareBE3} T. Feng, C. Ren, Y. Tian, M. Luo, H. Shi, J. Chen, and X. Zhou, \PRA \textbf{102}, 032220 (2020).
\bibitem{shareBT6} P. J. Brown, and R. Colbeck, \PRL \textbf{125}, 090401 (2020).

% 21-30
\bibitem{shareBT7} T. Zhang, and S. M. Fei, \PRA \textbf{103}, 032216 (2021).
\bibitem{shareBT8} S. Cheng, L. Liu, T. J. Baker, and M. J. W. Hall, \PRA \textbf{104}, L060201 (2021).
\bibitem{shareBT9} S. Saha, D. Das, S. Sasmal, D. Sarkar, K. Mukherjee, A. Roy, and S. S. Bhattacharya, \QIP \textbf{18}, 42 (2019).
\bibitem{shareST1} S. Sasmal, D. Das, S. Mal, and A. S. Majumdar, \PRA \textbf{98}, 012305 (2018).
\bibitem{shareST2} A. Shenoy H., S. Designolle, F. Hirsch, R. Silva, N. Gisin, and N. Brunner, \PRA \textbf{99}, 022317 (2019).
\bibitem{shareST3} D. Yao, and C. Ren, \PRA \textbf{103}, 052207 (2021).
\bibitem{shareST4} X. H. Han, Y. Xiao, H. C. Qu, R. H. He, X. Fan, T. Qian, and Y. J. Gu, \QIP \textbf{20}, 278 (2021).
\bibitem{shareET1} A. Bera, S. Mal, A. Sen(De), and U. Sen, \PRA \textbf{98}, 062304 (2018).
\bibitem{Ficek} Z. Ficek, and S. Swain, \textit{Quantum Interference and Coherence: Theory and Experiments} (Springer Series in Optical Sciences, Springer, Berlin, 2005).
\bibitem{coher} T. Baumgratz, M. Cramer, and M. B. Plenio, \PRL \textbf{113}, 140401 (2014).

%31-40
\bibitem{Plenio} A. Streltsov, G. Adesso, and M. B. Plenio, \RMP \textbf{89}, 041003 (2017).
\bibitem{Hu} M. L. Hu, X. Hu, J. C. Wang, Y. Peng, Y. R. Zhang, and H. Fan, \PR \textbf{762--764}, 1 (2018).
\bibitem{Wu} K. D. Wu, A. Streltsov, B. Regula, G. Y. Xiang, C. F. Li, and G. C. Guo, Adv. Quantum Technol. \textbf{4}, 2100040 (2021).
\bibitem{ad1} M. L. Hu, and H. Fan, \SCPMA \textbf{63}, 230322 (2020).
\bibitem{ad2} L. M. Zhang, T. Gao, and F. L. Yan, \SCPMA \textbf{64}, 260312 (2021).
\bibitem{ad3} Z. X. Jin, L. M. Yang, S. M. Fei, X. Li-Jost, Z. X. Wang, G. L. Long, and C. F. Qiao, \SCPMA \textbf{64}, 280311 (2021).
\bibitem{coen1} A. Streltsov, U. Singh, H. S. Dhar, M. N. Bera, and G. Adesso, \PRL \textbf{115}, 020403 (2015).
\bibitem{coen2} X. Qi, T. Gao, and F. Yan, \JPA \textbf{50}, 285301 (2017).
\bibitem{coen3} K. C. Tan, H. Kwon, C. Y. Park, and H. Jeong, \PRA \textbf{94}, 022329 (2016).
\bibitem{coqd1} Y. Yao, X. Xiao, L. Ge, and C. P. Sun, \PRA \textbf{92}, 022112 (2015).
\bibitem{coqd2} M. L. Hu, and H. Fan, \PRA \textbf{95}, 052106 (2017).
\bibitem{coqd3} X. Hu, and H. Fan, \SR \textbf{6}, 34380 (2016).
\bibitem{coqd4} X. Hu, A. Milne, B. Zhang, and H. Fan, \SR \textbf{6}, 19365 (2015).

% 41-50
\bibitem{naqc1} D. Mondal, T. Pramanik, and A. K. Pati, \PRA \textbf{95}, 010301 (2017).
\bibitem{naqc2} M. L. Hu, and H. Fan, \PRA \textbf{98}, 022312 (2018).
\bibitem{CHSH} J. F. Clauser, M. A. Horne, A. Shimony, and R. A. Holt, \PRL \textbf{23}, 880 (1969).
\bibitem{naqc3} M. L. Hu, X. M. Wang, and H. Fan, \PRA \textbf{98}, 032317 (2018).
\bibitem{sharenaqc} S. Datta, and A. S. Majumdar, \PRA \textbf{98}, 042311 (2018).
\bibitem{sharenaqce} S. Datta, and A. S. Majumdar, \PRA \textbf{99}, 019902 (2019).
\bibitem{von} J. von Neumann, \textit{Mathematical Foundations of Quantum Mechanics} (Princeton University Press, Princeton, 2018).
\bibitem{MUB1} W. K. Wooters, Found. Phys. 16, \textbf{391} (1986).
\bibitem{MUB2} W. K. Wooters, and B. D. Fields, Ann. Phys. \textbf{191}, 363 (1989).
\bibitem{weak1} Y. Aharonov, D. Z. Albert, and L. Vaidman, \PRL \textbf{60}, 1351 (1988).

% 51-59
\bibitem{weak2} I. M. Duck, P. M. Stevenson, and E. C. G. Sudarshan, \PRD \textbf{40}, 2112 (1989).
\bibitem{new1} X. D. Yu, D. J. Zhang, G. F. Xu, and D. M. Tong, \PRA \textbf{94}, 060302 (2016).
\bibitem{new2} C. L. Liu, X. D. Yu, and D. M. Tong, \PRA \textbf{99}, 042322 (2019).
\bibitem{Luders} P. Busch, \PRD \textbf{33}, 2253 (1986).
\bibitem{adv1} X. M. Hu, J. S. Chen, B. H. Liu, Y. Guo, Y. F. Huang, Z. Q. Zhou, Y. J. Han, C. F. Li, and G. C. Guo, \PRL \textbf{117}, 170403 (2016).
\bibitem{adv2} X. M. Hu, Y. Guo, B. H. Liu, Y. F. Huang, C. F. Li, and G. C. Guo, Sci. Adv. \textbf{4}, eaat9304 (2018).
\bibitem{adv3} X. M. Hu, W. B. Xing, B. H. Liu, Y. F. Huang, C. F. Li, G. C. Guo, P. Erker, and M. Huber, \PRL \textbf{125}, 090503 (2020).
\bibitem{adv4} X. M. Hu, C. Zhang, B. H. Liu, Y. Cai, X. J. Ye, Y. Guo, W. B. Xing, C. X. Huang, Y. F. Huang, C. F. Li, and G. C. Guo, \PRL \textbf{125}, 230501 (2020).
\bibitem{adv5} X. M. Hu, W. B. Xing, B. H. Liu, D. Y. He, H. Cao, Y. Guo, C. Zhang, H. Zhang, Y. F. Huang, C. F. Li, and G. C. Guo, Optica \textbf{7}, 738 (2020).



\end{thebibliography}
%\Non-BibTeX users please use

\end{document}